\documentclass{PoS}

\newcommand{\msbar}{\overline{\mbox{{\sc ms}}}}

\newcommand{\MOM}{\mbox{{\sc mom}}}

\newcommand{\lwrsim}{\raise0.3ex\hbox{$<$\kern-0.75em\raise-1.1ex\hbox{$\sim$}}}
\def\krto{ {\,\,\lower .8ex\hbox {$\longrightarrow \atop k \rightarrow 0$}\,\,}}

\def\bea{\begin{eqnarray} }
\def\beq{\begin{eqnarray} }

\def\eea{\end{eqnarray}}
\def\eeq{\end{eqnarray}}

\def\eq#1{Eq.~(\ref{#1})}

\title{Testing OPE for ghosts, gluons and $\alpha_s$}

\ShortTitle{OPE: ghosts, gluons and $\alpha_s$}

\author{\speaker{Konstantin Petrov}\thanks{P2IO LabEX}\\
        Laboratoire de l'Acc\'el\'erateur Lin\'eaire,
Centre Scientifique d'Orsay
B\^atiment 200 - BP 99
91898 ORSAY C\'edex \\
        E-mail: \email{petrov@lal.in2p3.fr}}
\author{ B.~Blossier, O.~P\`ene, Ph.~Boucaud\\Laboratoire de Physique Th\'eorique, 
Universit\'e de Paris XI; B\^atiment 210, 91405 Orsay Cedex; France}
\author{M.~Brinet\\LPSC, CNRS/IN2P3/UJF; 
53, avenue des Martyrs, 38026 Grenoble, France}
\author{F.~De Soto\\Dpto. Sistemas F\'isicos, Qu\'imicos y Naturales, 
Univ. Pablo de Olavide, 41013 Sevilla, Spain}

\author{V.~Morenas\\Laboratoire de Physique Corpusculaire, Universit\'e Blaise Pascal, CNRS/IN2P3 
63177 Aubi\`ere Cedex, France}

\author{J.~Rodr\'{\i}guez-Quintero\\Dpto. F\'isica Aplicada, Fac. Ciencias Experimentales; 
Universidad de Huelva, 21071 Huelva; Spain.}

\abstract{We present here our results on extracting Wilson coefficients from different quantities such as ghost and gluon propagators which are calculated by means of Lattice QCD. The results confirm the validity of our method for the calculation of the strong coupling constant as well as allow to estimate the range of momenta where OPE is applicable. }

\FullConference{Xth Quark Confinement and the Hadron Spectrum,\\
		October 8-12, 2012\\
		TUM Campus Garching, Munich, Germany}

\begin{document}

\section{Introduction}
It can be argued that the strong coupling constant $\alpha_s$, or, equivalently, the QCD scale $\Lambda_{QCD}$ are the fundamental quantities which require a high level of precision as they enter in most calculations, both perturbative and non-perturbative. They can be extracted both from experiment or numerically, with the help of Lattice QCD. The latter is a complicated procedure and it is imperative to understand the underlying mechanics to estimate correctly the systematic errors. For the review of such calculations see, for instance,  \cite{Alles:1996ka,Boucaud:1998bq,Becirevic:1999uc,Becirevic:1999hj,vonSmekal:1997is,Boucaud:2005xn,Boucaud:2008gn}. We presented our most recent results in~\cite{Boucaud:2011ug} and here we will demonstrate that our approach and estimation of systematics is justified. 

The work is based on ghost and gluon propagators, which we recently computed on the gauge configurations produced by the ETM Collaboration. These are the first to incorporate the effects of dynamical charm quarks. In other words, this is an ensemble of sample gauge fields  including $N_f$=2+1+1, two light degenerate and two heavy, dynamical quark flavours (see~\cite{Baron:2010bv,Baron:2011sf} for details of the simulations setup) with a twisted-mass fermion action~\cite{Frezzotti:2000nk,Frezzotti:2003xj}. Combining the propagators in Taylor scheme we were able to extract the strong coupling constant, which agrees pretty well with the experimental 
results from $\tau$ decays~\cite{Bethke:2011tr} and "{\it world average}" from PDG~\cite{Nakamura:2010zzi} at all scales from $\tau$ to $Z_0$. At the same time it became clear that  invocation of nonperturbative OPE corrections is unequivocally needed to account properly for the lattice data of the coupling in Taylor scheme. 
Within the OPE procedure we expand matrix elements of any non-local operator in terms of local operators, organized by their momentum dimensions. Using the sum rules factorization~\cite{Shifman:1978bx,Shifman:1978by}, these can be expressed as a coefficient to be computed in perturbation (Wilson coefficient) and the nonperturbative condensate of a local operator. For such procedure to be consistent it is required that (up to a proper renormalization constant for the local operator) the value of the condensate should be the same for any Green Function.

In refs.~\cite{Blossier:2011tf,Blossier:2012ef} for $N_f$=2+1+1 and in refs.~\cite{Boucaud:2008gn,Blossier:2010ky} for $N_f$=0 and $N_f=$2 quark flavours respectively, we calculate the running coupling by equating numerical results and the OPE prediction. The OPE is clearly dominated by the landau gauge gluon condensate $A^2$ which has been elsewhere very much studied ({\it e.g.}, see \cite{Boucaud:2000nd,Gubarev:2000nz,Kondo:2001nq,Verschelde:2001ia,Dudal:2002pq,RuizArriola:2004en,Vercauteren:2011ze}). It can be noted that to calculate the coupling more reliably we combined the bare lattice propagators  in such a manner that the cut-off dependence is minimal, which  prevents the check for the universality of condensates, as gluon and ghost propagators are not to be independently analyzed. However our previous studies in quenched approximation suggested that such universality (even including the three-gluon vertex) is indeed compatible with the numerical data ($N_f$=0)~\cite{Boucaud:2000nd,Boucaud:2001st,DeSoto:2001qx,Boucaud:2005xn,Boucaud:2005rm,Boucaud:2008gn}. This has been also supported by the calculations with $N_f$=2 dynamical flavours~\cite{Blossier:2010ky,Blossier:2010vt}. Now we will address this question in the $N_f=$2+1+1 case, to clarify the nature itself of the unavoidable nonperturbative corrections and the involved condensates.

 \section{Basics and Simulations}
\label{sec:OPE}

The Taylor coupling can be obtained directly from the gluon and ghost dressing functions it follows~\cite{Boucaud:2008gn},
%-------------
\beq\label{eq:alphaTdef}
\alpha_T(p^2) \ = \ \frac{g_0^2(a^{-1})}{4\pi} \ F^2(p^2,a^{-1}) \ G(p^2,a^{-1})  \ .
\eeq
%-------------
Substituting the dressing functions by their OPE expressions we are left with 
%-------------
\beq\label{eq:alphaTNP}
\alpha_T(p^2) \ = \ \alpha_T^{\rm pert}(p^2) \left( 1 +  9 \ 
R\left(\alpha_T^{\rm pert}(p^2),\alpha_T^{\rm pert}(\mu^2)\right) 
\left(\frac{\alpha(p^2)}{\alpha(\mu^2)}\right)^{\frac{27}{100}} \ \frac{g^2(\mu^2) \langle A^2 \rangle_{{\rm R},\mu^2}}{4\left(N^2_c-1\right) p^2} \ + \ \cdots \right)
\eeq
%-------------
where $R$ is known to four loops.

%-------------
%\beq\label{eq:Ra0}
%R\left(\alpha,\alpha_0\right) \ = \ \left( 1 + 1.03735 \ \alpha + 1.07203 \ \alpha^2 + 1.5%9654 \ \alpha^3 \right) \left(1 - 0.54993 \ \alpha_0 - 0.14352 \ \alpha_0^2 -  0.07339 \ \alpha_0^3 \right) \ .
%\eeq
%------------
We refer to the upcoming paper \cite{upcoming} for the details of this calculation.

The ghost and gluon propagators are defined as follows:
 
%---------------
\beq
 \Delta^{ab}_{\mu\nu}(q)&=&\left\langle A_{\mu}^{a}(q)A_{\nu}^{b}(-q)\right\rangle
 = \delta^{ab}\left(\delta_{\mu\nu}-\frac{q_\mu q_\nu}{q^2}\right) \frac{G(q^2)}{q^ 2} \;,
\nonumber \\
F^{ab}(q^2) &=& \frac 1 V \ \left\langle \sum_{x,y}
 \exp[iq\cdot(x-y)] \left( M^{-1} \right)^{ab}_{xy} \right\rangle
 =\delta^{ab}\frac{F(q^2)}{q^2} \;;
\label{green}
\eeq
%---------------
where $A_\mu^ a$ is the gauge field and $M^{ab}\equiv \partial_\mu D^{ab}_\mu$ is the Fadeev-Popov operator.The lattice setup is described in \cite{Blossier:2012ef} and summarized in Tab.~\ref{tab:setup}. The $O(4)$-breaking lattice artefacts have been cured by the so-called $H(4)$-extrapolation procedure~\cite{Becirevic:1999uc,deSoto:2007ht}. These estimates for bare gluon and ghost dressing functions appear plotted in Fig.~\ref{fig:green}. 

%----------------------------
\begin{table}[ht]
\begin{center}
\begin{tabular}{|c|c|c|c|c|c|c|}
\hline
$\beta$ & $\kappa_{\rm crit}$ & $a \mu_l$ & $a \mu_\sigma$ & $a \mu_\delta$ &
$(L/a)^3\times T/a$ & confs. \\
\hline
1.90 & 0.1632700 & 0.0040 & 0.150 & 0.1900 & $32^3\times 64$ & 50 \\	
1.95 & 0.1612400 & 0.0035 & 0.135 & 0.1700 & $32^3\times 64$ & 50 \\
2.10 &  0.1563570 & 0.0020 & 0.120 & 0.1385 & $48^3\times 96$ & 100 \\  
\hline
\end{tabular}
\caption{Lattice setup parameters for the ensembles we used in this paper: $\kappa_{\rm crit}$ is the critical value for the standard hopping parameter for the bare untwisted mass; $\mu_l$
stands for the twisted mass for the two degenerated light quarks,
while $\mu_\sigma$ and $\mu_\delta$ define the heavy quarks
twisted masses; the last column indicates the number of gauge
field configurations exploited. This implies that the strange quark mass is roughly set to 95 MeV and 
the charm one to 1.51 GeV (in $\msbar$ at 2 GeV), while degenerate light quark masses range from 20 to 50 MeV (The lightest pseudoscalar masses approximately range from 270 to 510 MeV).}
\label{tab:setup}
\end{center}
\end{table}
%------------------------------

\section{Testing the OPE}
\label{subsec:remove}

\begin{figure}[th]
 \begin{center}
	\begin{tabular}{cc}
    \includegraphics[width=6.5cm]{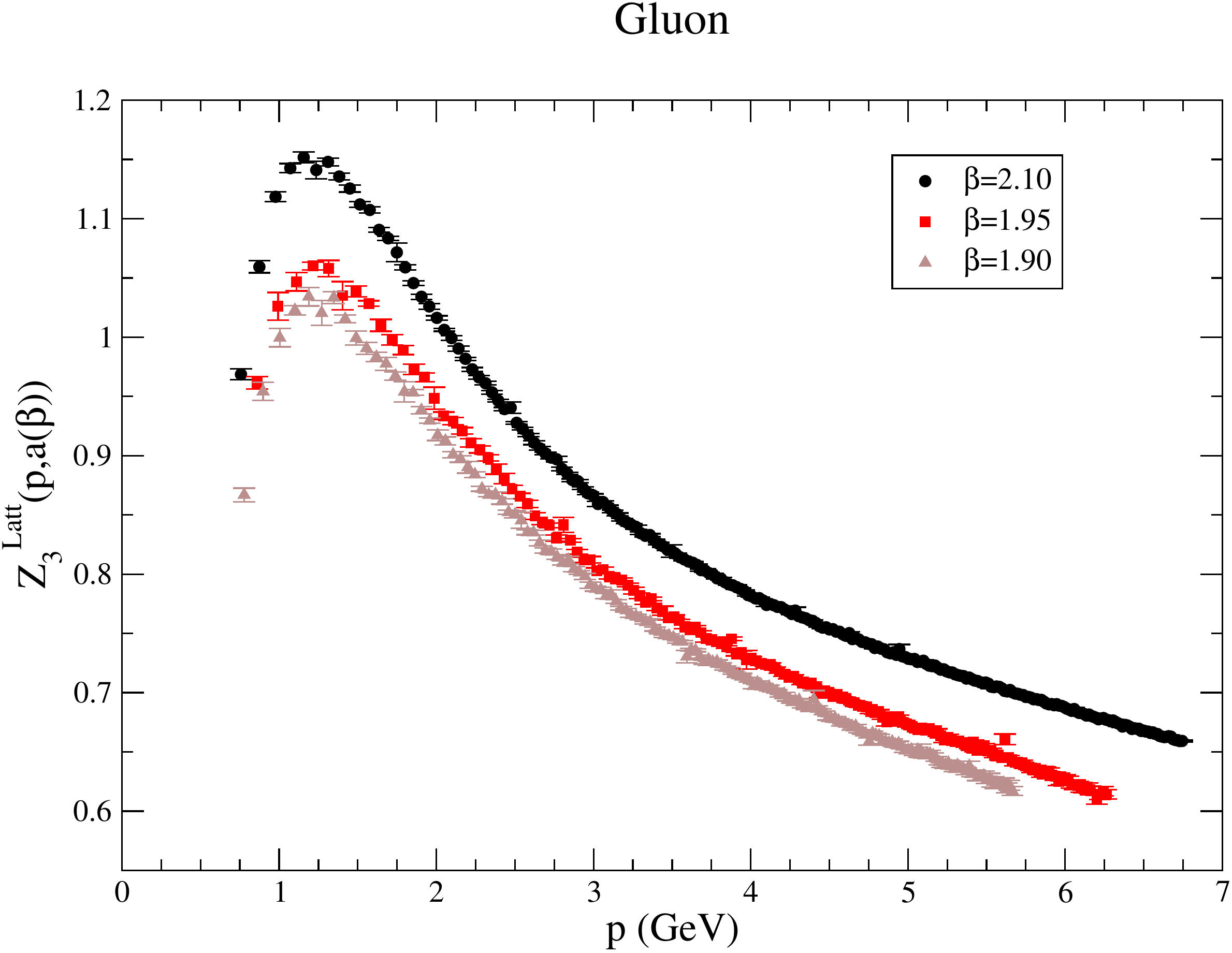} & 
    \includegraphics[width=6.5cm]{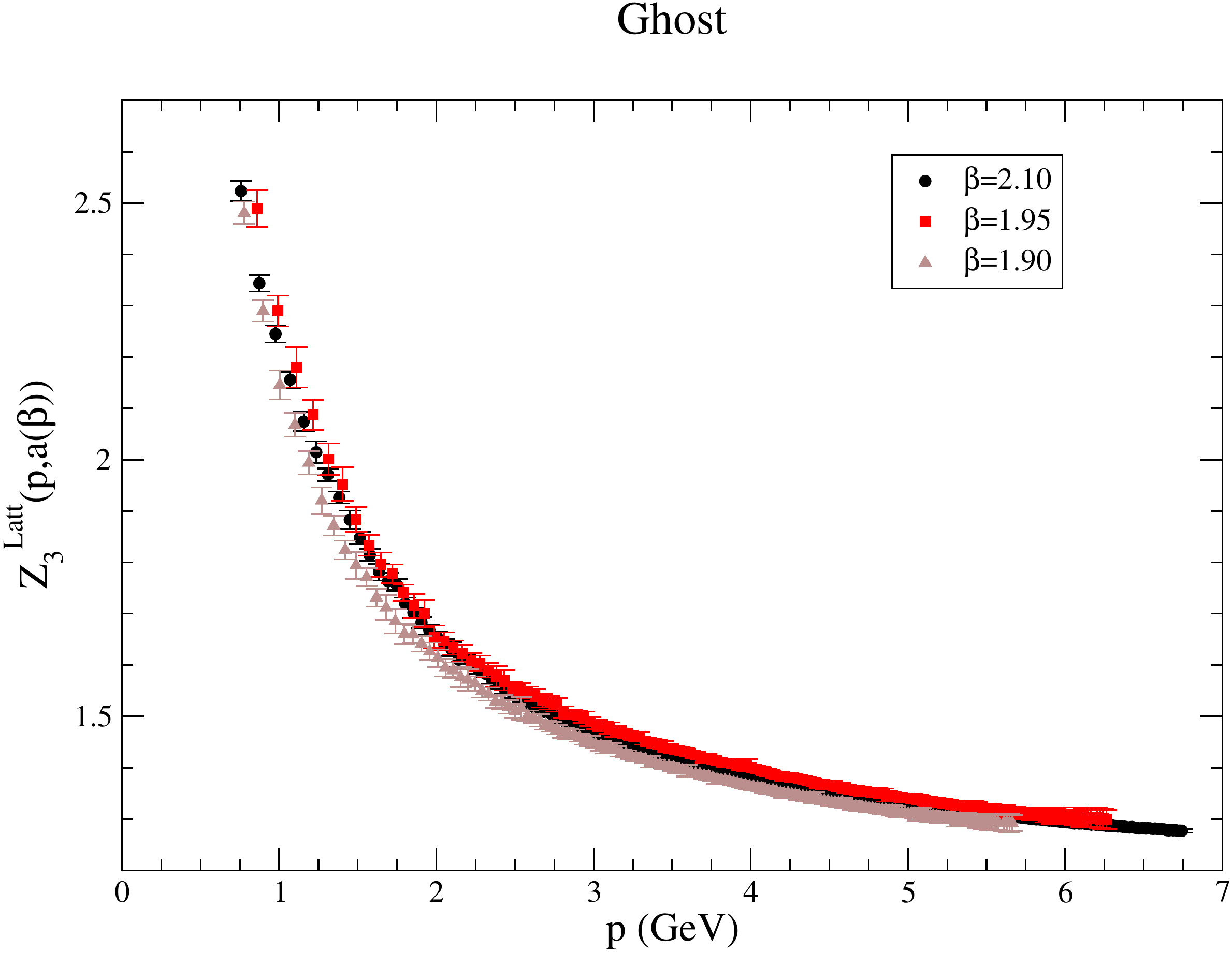} 
    \\
     \includegraphics[width=6.5cm]{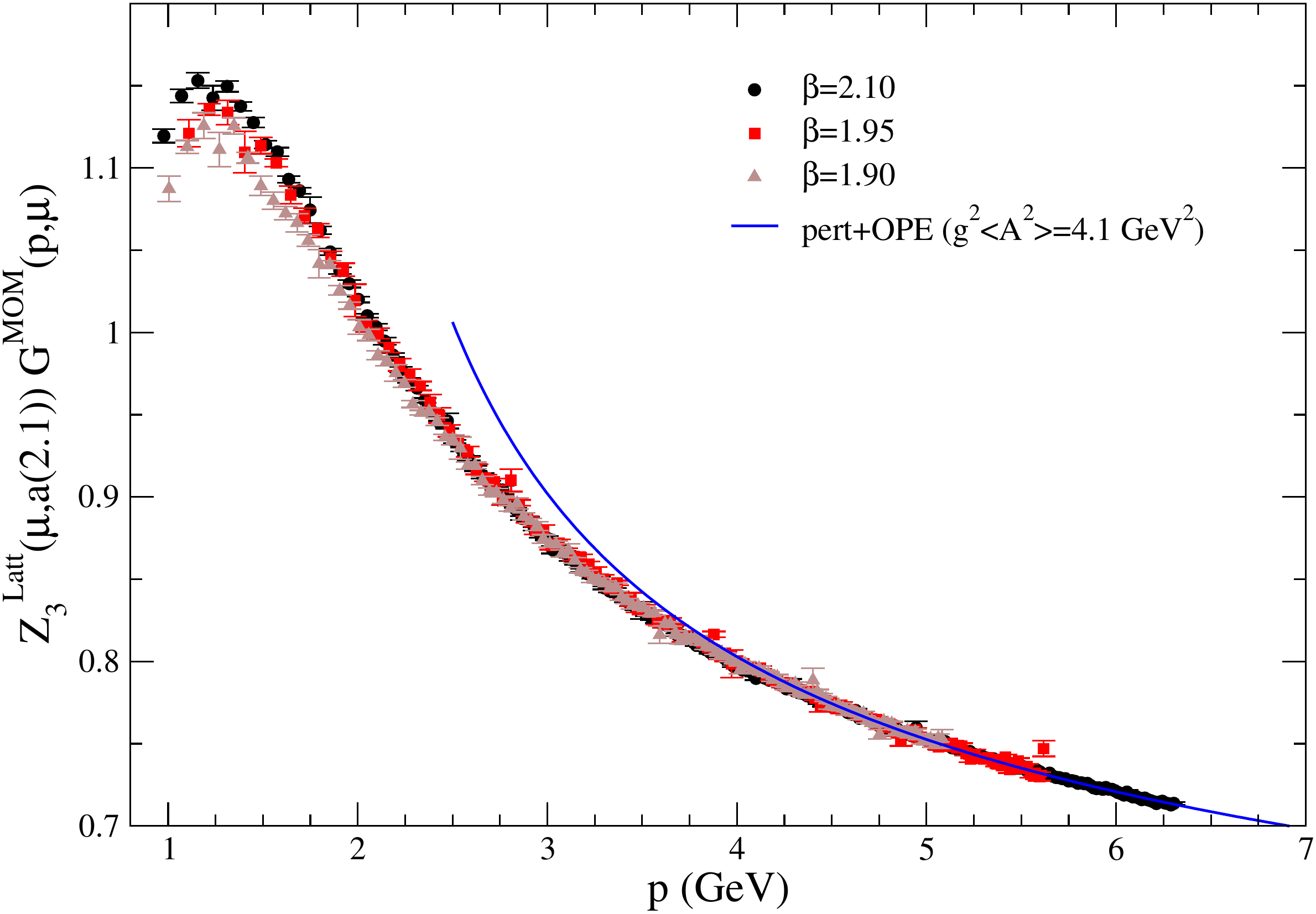} & 
    \includegraphics[width=6.5cm]{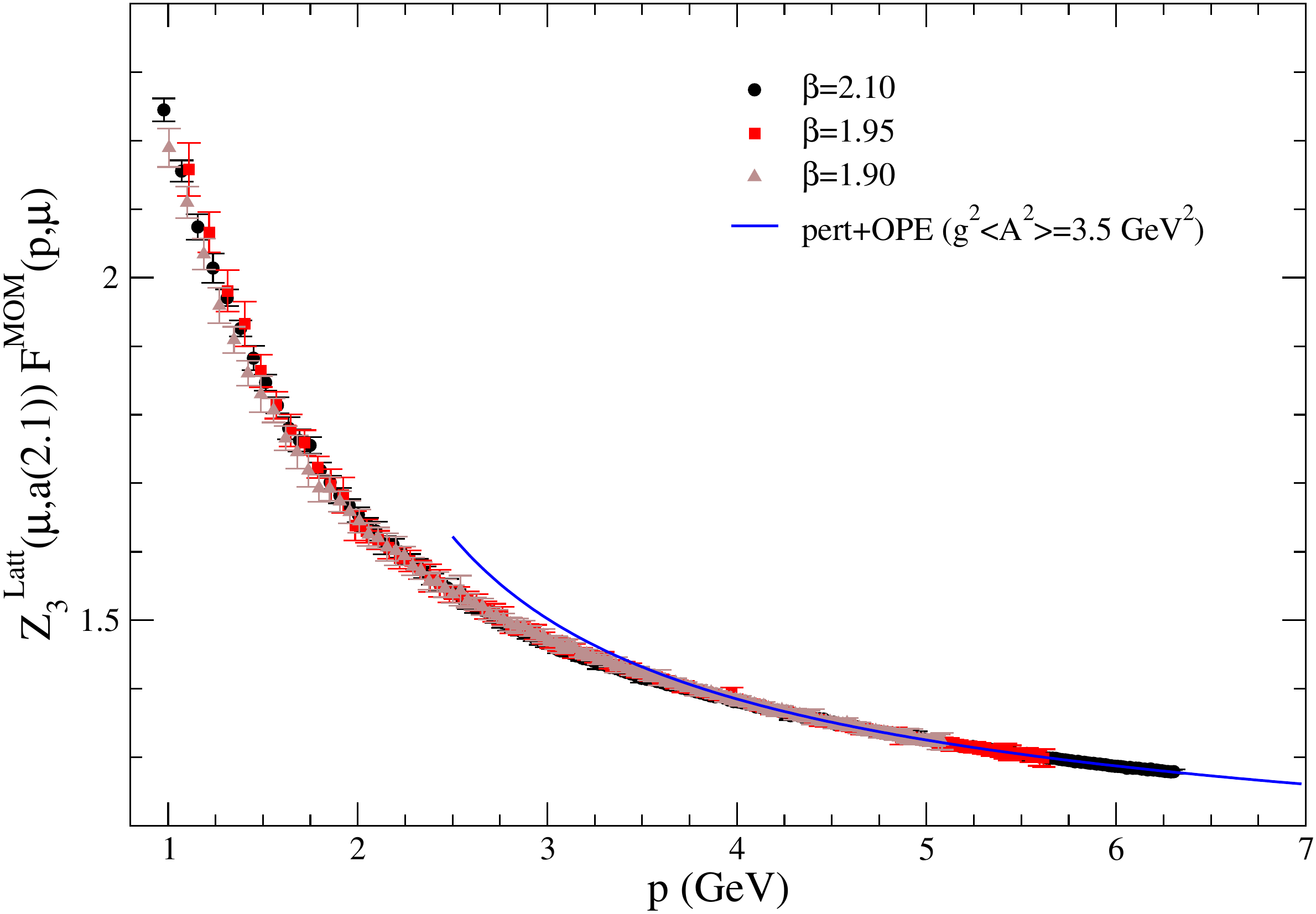} 
   
	\end{tabular}
  \end{center}
\caption{\small Bare gluon and ghost propagators as they result from lattice computations before (top) and 
after being cured of $O(4)$-invariant lattice artefacts and properly rescaled to appear superimposed (bottom). The blue solid line correspond to the best fit with the OPE formula.
}
\label{fig:green}
\end{figure}
%----------------------------------------

The gluon and ghost propagator lattice data shown in the upper plots of Fig.~\ref{fig:green} are still affected by $O(4)$-invariant lattice artefacts. As we did in ref.~\cite{Blossier:2011tf,Blossier:2012ef,Blossier:2010vt} for the
Taylor coupling and the vector quark propagator, the gluon propagator renormalization constant $Z_3^{\rm Latt}$ in MOM scheme can be written as 
%-------------
\beq\label{eq:Z3g}
Z_3^{\rm Latt}(p^2,a^{-1}) \ = \ G(p^2,a^{-1}) + c_{a2p2} \ a^2 p^2 \ ,
\eeq
%-------------
which corresponds to the lattice gluon dressing function. In r.h.s., the artefacts-free gluon dressing is noted as $G$. It can be shown that we can rewrite \eq{eq:Z3g} as:
%-------------
\beq\label{eq:art}
\frac{Z_3^{\rm Latt}(p^2,a^{-1})}{G^{\rm NP}\left(\frac{p^2}{\mu^ 2},\alpha(\mu^ 2)\right)} \ = \
z(\mu^2,a^{-1}) \ + \ c_{a2p2} \ \frac{a^2 p^2}{G^{\rm NP}\left(\frac{p^2}{\mu^ 2},\alpha(\mu^ 2)\right)}
\eeq
%-------------
with 
{\small
\beq
\label{eq:GNP}
G^{\rm NP}\left(\frac{p^2}{\mu^ 2},\alpha(\mu^ 2)\right) \ = \ 
G_0^{\MOM}\left(\frac{p^2}{\mu^2},\alpha(\mu^2)\right)
\left( 1 +  
%\frac{c_2^{\wmsbar}\left(\frac{p^2}{\mu^2},\alpha(\mu^2)\right)}
%{c_0^{\MOM}\left(\frac{p^2}{\mu^2},\alpha(\mu^2)\right)}
%\frac{c_2^{\wmsbar}\left(\frac{p^2}{\mu^2},\alpha(\mu^2)\right)}
%{c_0^{\MOM}\left(\frac{p^2}{\mu^2},\alpha(\mu^2)\right)}
C_w\left(\frac{p^2}{\mu^2},\alpha(\mu^2)\right) \ \frac{\langle A^2 \rangle_{{\rm R},\mu^2}}{4\left(N^2_c-1\right) p^2} 
\ + \ \cdots \ \right), 
\eeq 
}
%--------------
where Wilson coefficient $C_w\left(\frac{p^2}{\mu^2},\alpha(\mu^2)\right)$, in the appropriate renormalization scheme, is known 
at the four-loop order with the help of the perturbative expansions for the 
gluon propagator MOM anomalous dimension and Taylor scheme beta function~\cite{Chetyrkin:2000dq}. 
 $G^{\rm NP}$ is expected to describe properly the nonperturbative running of the gluon dressing from lower bound of 
 $p=4.5$ GeV, below which the lattice data deviate from the nonperturbative prediction only including the OPE leading power correction, to the upper bound of  $a^2p^2 \sim 3.5$, above where the impact of higher-order lattice artefacts cannot be neglected any longer.
Now we can perform a linear fit and determine the overall factor, $z(\mu,a^{-1})$, for each data set
and the coefficients $c_{\rm a2p2}$, which should be almost the same as they depend only logarithmically on the lattice spacing $a(\beta)$ by dimensional arguments (see tab.~\ref{tab:res}).
This way we can remove the $O(4)$-invariant artefacts away from the lattice data for 
all available momenta in the aforementioned range, by applying \eq{eq:Z3g}, to be left with the continuum nonperturbative running for the gluon dressing; 
while the overall factors $z(\mu,a^{-1})$ can be used to rescale the data obtained at different $\beta$ such that 
they all will follow the same curve (see down plots of fig.~\ref{fig:green}).

As $G^{\rm NP}$ depends on $\Lambda_{\overline{\rm MS}}$ and the gluon condensate, 
$g^2\langle A^2 \rangle$, their values have to be known prior to the removal of $O(4)$-invariant artefacts. 
In practice, we will take $\Lambda_{\overline{\rm MS}}$ to be known from $\tau$ decays 
($\alpha_{\msbar}(m_\tau^2)=0.334(14)$~\cite{Bethke:2011tr}, {\it i.e.} $\Lambda_{\overline{\rm MS}}=311$ MeV for $N_f=4$) 
and will search for a value of $g^2\langle A^2 \rangle$ such that one gets the best matching of the rescaled gluon dressings 
from the three data sets. The details and the intermediate results of this procedure will be explained in \cite{upcoming}. Here we just present the  results  in Tab.~\ref{tab:comp}.

{\small

\begin{table}[t]
\small
\begin{center}
\begin{tabular}{|c|c|c|}
\hline
& gluon & ghost \\
\hline
$g^2 \langle A^2 \rangle$ [GeV$^2$] & 4.7 & 3.2 \\
\hline
\begin{tabular}{c}
\rule[0cm]{0.95cm}{0cm}
$\beta$  
\rule[0cm]{0.95cm}{0cm}
\\
\hline
$z(\mu^2,a^{-1})$ 
 \\
\hline
$c_{a2p2}$
\end{tabular}
&
\begin{tabular}{c|c|c}
 2.10 & 1.95 & 1.90 \\
\hline
 0.6495(5) & 0.6068(26) & 0.5971(30) \\
 \hline
 -0.0097(2) & -0.0071(8) & -0.0080(8)
\end{tabular}
&
\begin{tabular}{c|c|c}
 2.10 & 1.95 & 1.90 \\
\hline
 1.215(6) & 1.229(20) & 1.193(31) \\
\hline
0.0002(19) & -0.0012(44) & 0.0011(69) 
\end{tabular}
\\ 
\hline
\hline
$g^2 \langle A^2 \rangle$ [GeV$^2$] & 4.1 & 4.1 \\
\hline
\begin{tabular}{c}
$z(\mu^2,a^{-1})$ 
 \\
\hline
\rule[0cm]{0.75cm}{0cm}
$c_{a2p2}$
\rule[0cm]{0.75cm}{0cm}
\end{tabular}
&
\begin{tabular}{c|c|c}
 0.6531(5) & 0.6106(26) & 0.6014(30) \\
 \hline
 -0.0104(2) & -0.0077(8) & -0.0086(8)
\end{tabular}
&
\begin{tabular}{c|c|c}
 1.206(6) & 1.218(19) & 1.182(30) \\
\hline
0.0022(19) & 0.0004(44) & 0.003(7) 
\end{tabular}
\\ 
\hline
\end{tabular}
\end{center}
\caption{\small The overall factors, $z(\mu^2,a^{-1})$, and the coefficients $c_{a2p2}$, for optimal values of $g^2\langle A^2 \rangle$. As discussed there, $c_{a2p2}$ is thought not to depend very much on $\beta$ and so is fairly obtained from the results shown here. The renormalization point, $\mu$, for the overall factor is fixed at 10 GeV. Statistical errors have been computed by applying the Jackknife method.}
\label{tab:res}
\end{table}

\begin{table}[ht]
\begin{center}
\begin{tabular}{|c|c|c|c|c|}
\hline
& gluon & ghost & $\alpha_T$ \cite{Blossier:2011tf} & $\alpha_T$ \cite{Blossier:2012ef} 
\\
\hline 
$g^2 \langle A^2 \rangle$ [GeV$^2$] & 4.7(1.6) & 3.1(1.1) & 4.5(4) & 3.8(1.0) 
\\
\hline
\end{tabular}
\end{center}
\caption{\small Results for $g^2\langle A^2 \rangle$ independently obtained from gluon and ghost propagators in this paper and from the Taylor coupling analysis in refs.~\cite{Blossier:2011tf,Blossier:2012ef}.}
\label{tab:comp}
\end{table}
}

Finally, one can impose the condensates to be the same for both gluon and ghost propagators, which leads to $g^2\langle A^2 \rangle=4.1$ GeV$^2$, that appears to be again in a pretty good agreement with the previous results in refs.~\cite{Blossier:2011tf,Blossier:2012ef}. In the following, we will use the artefacts-free gluon and ghost dressing functions, obtained with this last value for the gluon condensate, shown in Fig.~\ref{fig:green}.

Now we can directly test the validity of the OPE to describe consistently and accurately different Green functions by confronting the directly extracted numbers with theoretical predictions:
{\small
\beq\label{eq:W1}
p^2 \ \left[ \frac{G(p^2,a^{-1})}{z(\mu^2,a^{-1})} - G^{\rm MOM}_0\left(\frac{p^2}{\mu^2},\alpha(\mu^2)\right) \right] \ = \
G^{\rm MOM}_0\left(\frac{p^2}{\mu^2},\alpha(\mu^2)\right) C_w\left(\frac{p^2}{\mu^2},\alpha(\mu^2)\right) 
\frac{\langle A^2 \rangle_{{\rm R},\mu^2}}{4\left(N^2_c-1\right)}\ ,
\eeq
%-----------
}
where the l.h.s. is computed after the removal of the artefacts.

\begin{figure}[hbt]
\begin{center}
\begin{tabular}{cc}
	\includegraphics[width=6.5cm]{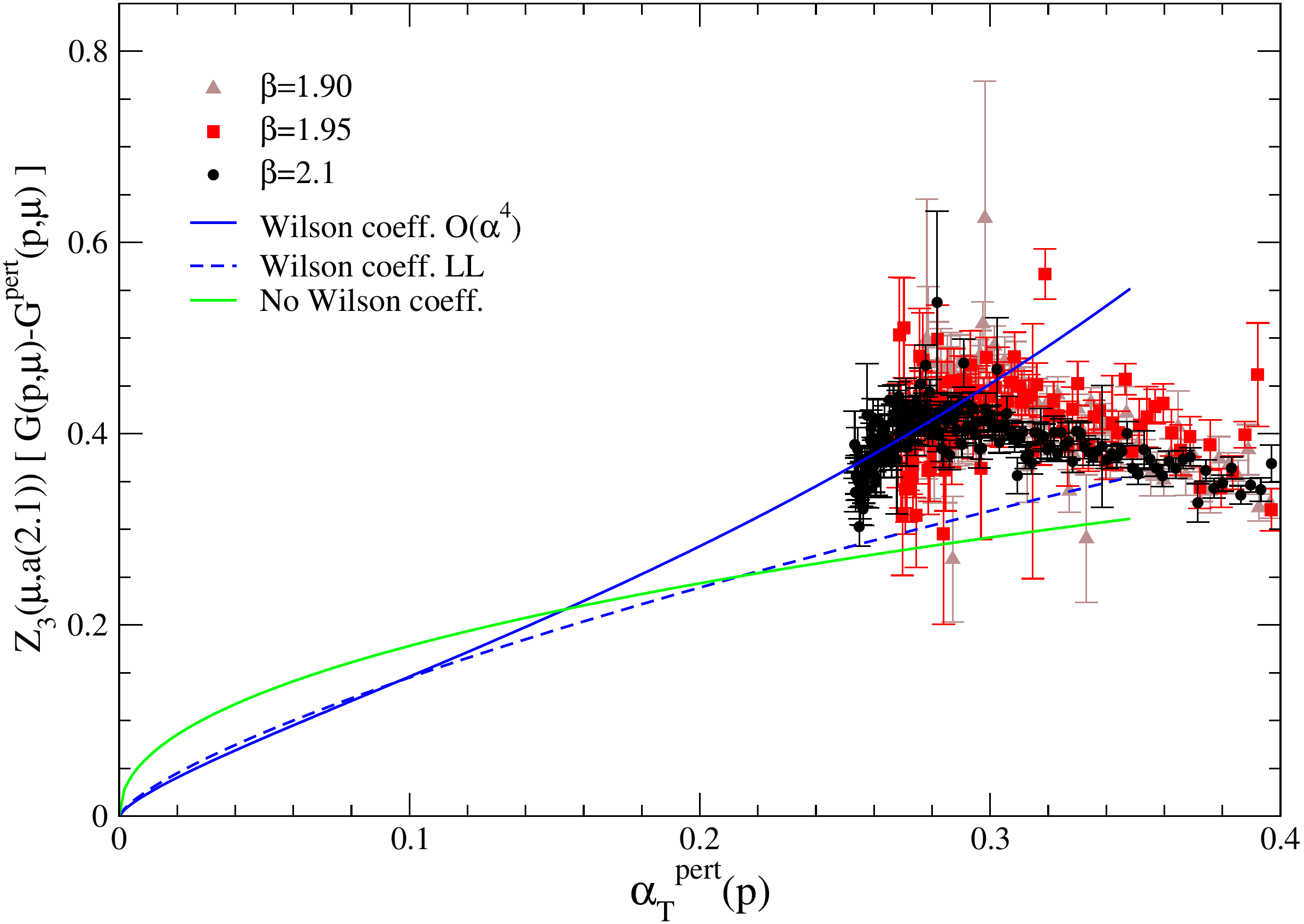}      
     & 
     \includegraphics[width=6.5cm]{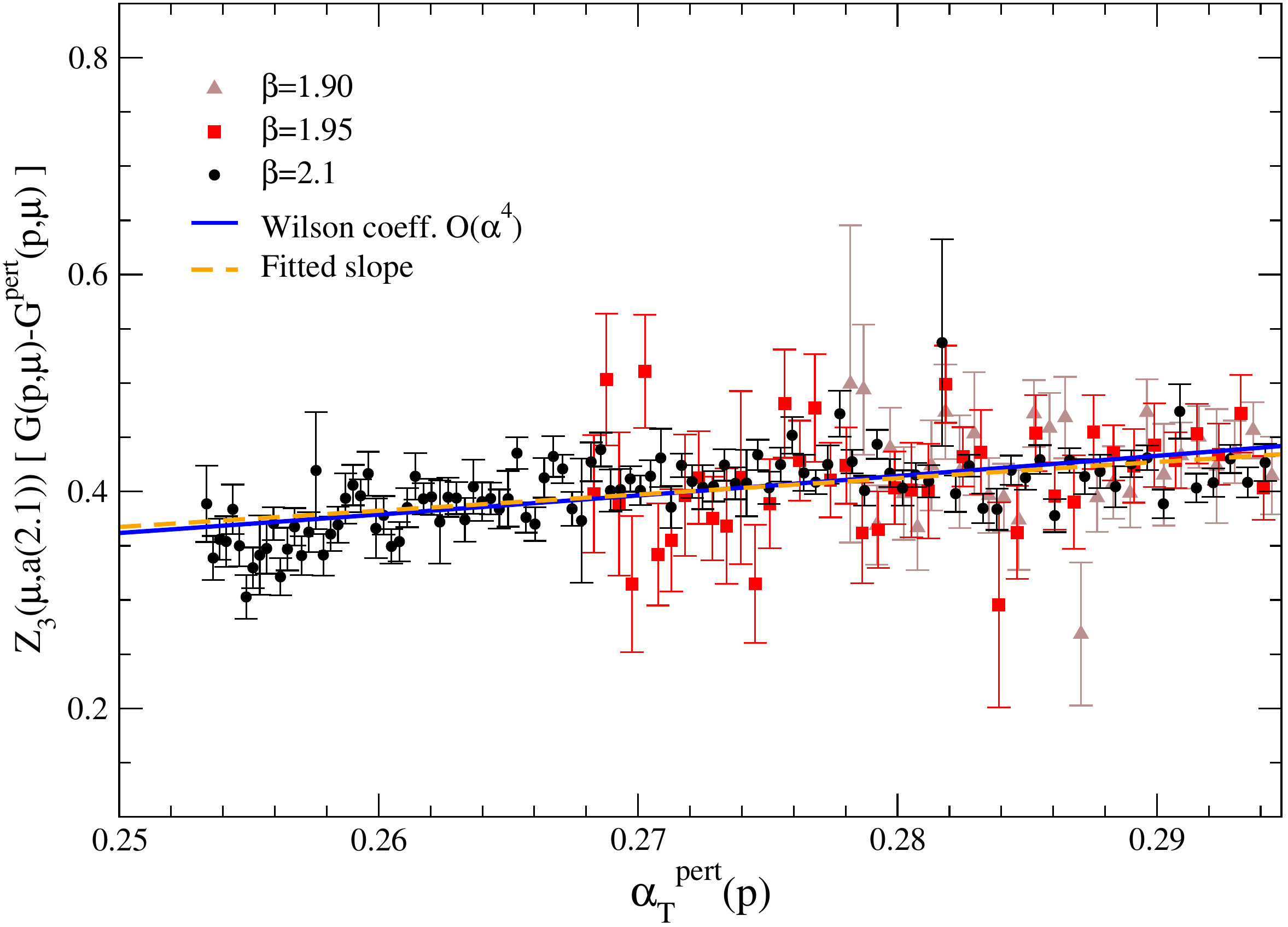} 
     \\
     	\includegraphics[width=6.5cm]{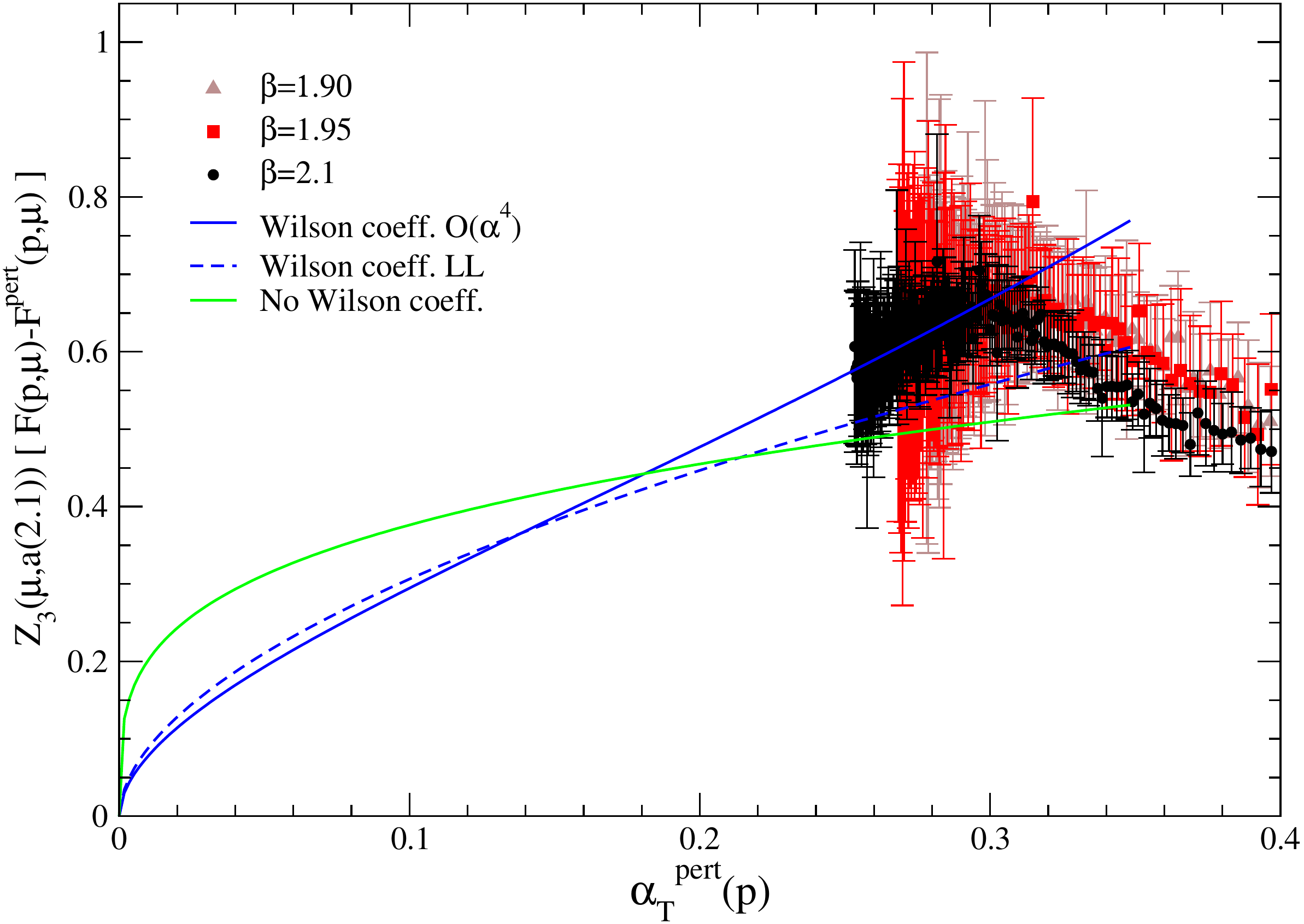}      
     & 
     \includegraphics[width=6.5cm]{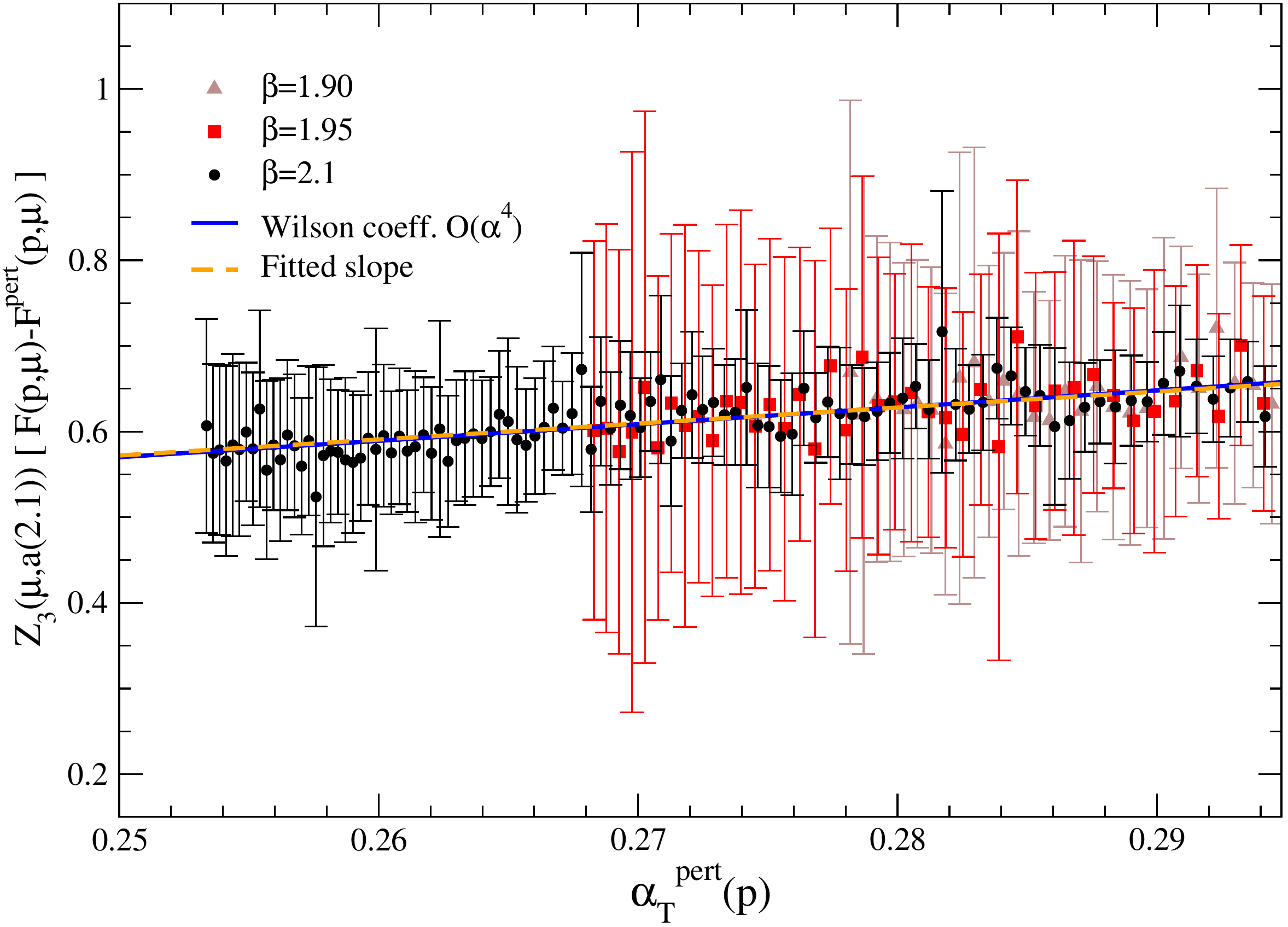} 
\end{tabular}
\end{center}
\caption{ The left plots shows the difference between the rescaled lattice data for the gluon (top) and ghost (bottom) dressing functions and their perturbative estimate, multiplied by the square of momentum, plotted versus the perturbative Taylor coupling at the same momentum. For perturbative evaluations, $\Lambda_{\msbar}$ is taken to be known from $\tau$ decays. Blue dotted lines correspond to using a leading logarithm approximation for Wilson coefficient, and green solid line describes the purely perturbative ansatz. The right plots zoom into the region, from 4.5 to 6.3 GeV, where the behavior, predicted by the OPE analysis, is observed. The dotted orange line shows a linear fit of data that shows a good agreement with the predicted running.}
\label{fig:Wilson}
\end{figure}

\begin{figure}[hbt]
\begin{center}
%\begin{tabular}{cc}
	\includegraphics[width=12cm]{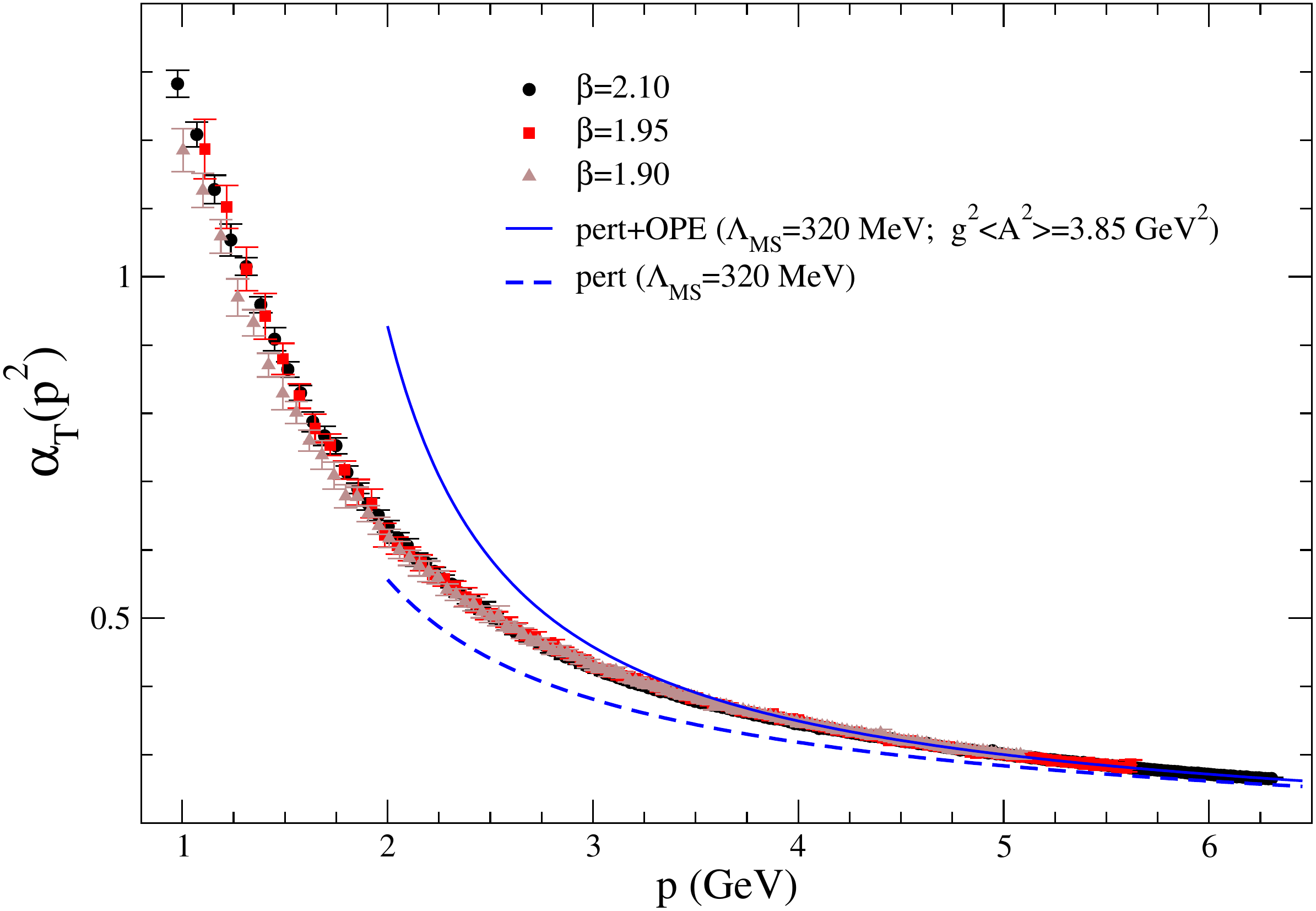}      
%\end{tabular}
\end{center}
\caption{\small The Taylor coupling computed with the artefact-free lattice dressing functions and the best-fit parameters for $\Lambda_{\msbar}$ and $g^2\langle A^2 \rangle$ (solid blue line). For the sake of comparison, the four loop perturbative prediction, evaluated for the same $\Lambda_{\msbar}$, is also plotted with a dashed blue line.
}
\label{fig:alpha} 
\end{figure}

 On \eq{eq:W1}'s r.h.s., the running with momenta can be easily expressed as a perturbative series in terms of the Taylor coupling, being proportional to the gluon condensate, $g^2 \langle A^2 \rangle$, which we already calculated.  As one can see from the Fig.~\ref{fig:Wilson}, the nonperturbative corrections are totally unavoidable for both the gluon (top) and ghost (bottom) cases.  Also, the Wilson coefficient derived at the order ${\cal O}(\alpha^4)$ in sec.~\ref{sec:OPE} gives a high-quality fit of data (blue solid lines in both plots). Had we not included the Wilson coefficient running ({i.e.}, dropped $C_w$ from \eq{eq:W1}'s r.h.s.), or just included its leading logarithm approximation, a flatter slope would be obtained, not describing properly the data (blue dotted and green solid lines). A similar situation can be observed in the case of Taylor coupling itself, see Fig.\ref{fig:alpha}.

\section*{Acknowledgments} 

We thank for the support the Spanish MICINN FPA2011-23781 and 
``Junta de Andalucia'' P07FQM02962 research projects, and  
the CC-IN2P3 (CNRS-Lyon), IDRIS (CNRS-Orsay), TGCC (Bruy\'eres-Le-Chatel) and CINES (Montpellier) as well as for GENCI support under Grant 052271. Speaker is supported by the P2IO LABEX initiative of France. 
\bibliography{total}
\bibliographystyle{unsrt} % not to be used with revtex

%\begin{thebibliography}{99}
%\bibitem{...} 
%....
%
%\end{thebibliography}

\end{document}